\documentclass[11pt]{article}
\usepackage{mathrsfs}
\usepackage[centertags]{amsmath}
\usepackage{amsfonts}
\usepackage{amssymb}
\usepackage{amsthm}
\usepackage{epsfig}
\usepackage{color}
\allowdisplaybreaks[4]
\usepackage{indentfirst}
\usepackage{amssymb,color}

\date{}

\newtheorem{theorem}{Theorem}[]
\newtheorem{proposition}{Proposition}[]

\newtheorem{lemma}{Lemma}[]

\numberwithin{equation}{section}

\def\P{\operatorname*{P}}
\def\C{\operatorname*{\mathbb{C}}}
\def\SN{\operatorname*{SN}}
\def\R{\operatorname*{\mathbb{R}}}
\begin{document}
\title{Rates of convergence of extremes from skew normal samples}
\author{{$^a$Xin Liao\quad $^a$Zuoxiang Peng  \quad $^b$Saralees Nadarajah \quad $^c$Xiaoqian Wang}\\
{\small\it $^a$School of Mathematics and Statistics, Southwest University, Chongqing, 400715  China}\\
{\small\it $^b$School of Mathematics, University of Manchester, Manchester, United Kingdom}\\
{\small\it $^c$Institute of Mathematics, School of Mathematics Sciences, Nanjing Normal University, Nanjing, 210097 China}}
\maketitle

\begin{quote}
{\bf Abstract}~~For a skew normal random sequence,
convergence rates of the distribution of its
partial maximum to the Gumbel extreme value distribution are derived.
The asymptotic expansion of the distribution of the normalized maximum is given under an optimal choice of norming constants.
We find that the optimal
convergence rate of the normalized maximum to
the Gumbel extreme value distribution is proportional to $1/\log n$.



\end{quote}

\section{Introduction}
\label{sec1}

The biggest weakness of the normal distribution is its inability to model skewed data.
This has led to several skewed extensions of the normal distribution.
The most popular and the most studied of these extensions is the skew normal distribution due to Azzalini (1985).
A random variable $X$ is said to have a standard skew-normal distribution with shape
parameter $\lambda\in R$ (written as $X \sim \SN(\lambda)$) if its
probability density function (pdf) is
\begin{eqnarray*}
f_{\lambda}(x)= 2\phi(x)\Phi(\lambda x),
\quad
-\infty < x < +\infty,
\end{eqnarray*}
where $\phi (\cdot)$ denotes the standard normal pdf
and $\Phi(\cdot)$ denotes the standard normal cumulative distribution function (cdf).
It is known that $\SN (0)$ is a standard normal random variable.

The skew normal distribution has received more applications than any other extension of the normal distribution.
Its applications are too many to list.
Some applications of the skew normal distribution that have appeared in the past year alone include:
the distribution of threshold voltage degradation in nanoscale
transistors by using reaction-diffusion and percolation theory (Islam and Alam, 2011);
population structure of Schima superba in Qingliangfeng National Nature Reserve (Liu et al., 2011);
rain height models to predict fading due to wet snow on terrestrial links (Paulson and Al-Mreri, 2011);
modeling of seasonal rainfall in Africa (Siebert and Ward, 2011);
modeling of HIV viral loads (Bandyopadhyay et al., 2012);
multisite flooding hazard assessment in the Upper Mississippi River (Ghizzoni et al., 2012);
modeling of diabetic macular Edema data (Mansourian et al., 2012);
risks of macroeconomic forecasts (Pinheiro and Esteves, 2012);
modeling of current account balance data (Saez et al., 2012);
automated neonatal EEG classification (Temko et al., 2012).

The aim of this note is to establish the convergence rate of
the distribution of the maxima for samples obeying $\SN(\lambda)$.
Chang and Genton (2007) showed that $\SN(\lambda)$ belongs to the domain of attraction of
the Gumbel extreme value distribution $\Lambda(x)=\exp(-\exp(-x))$.
Rates of convergence of the distribution of maxima for a
sequence of independent $SN (0)$ random variables were
studied by Hall (1979), Leadbetter et al. (1983) and Nair (1981).
Precisely speaking, Leadbetter et al. (1983) proved that
\begin{eqnarray*}
\Phi^{n}\left( \alpha_{n}x+\beta_{n} \right) - \Lambda(x) \sim
\frac {e^{-x}\exp \left( -e^{-x}\right)}{16}
\frac {\left(  \log \log n\right)^{2}}{\log n}
\end{eqnarray*}
holds for large $n$ with normalized constants $\alpha_{n}$ and
$\beta_{n}$  given by
\begin{eqnarray*}
\alpha_{n}=(2\log n)^{-\frac {1}{2}}
\quad
\mbox{and}
\quad
\beta_{n}=\alpha_{n}^{-1} - \frac {\alpha_{n}}{2}\left( \log \log n +\log 4\pi \right).
\end{eqnarray*}
The optimal uniform convergence rate of
$\Phi^{n}(\widetilde{a}_{n}x+\widetilde{b}_{n})$ to $\Lambda(x)$ due to Hall (1979) is
\begin{eqnarray*}
\frac {\C_{1}}{\log n}<\sup_{x\in \R}\left| \Phi^{n}\left(
\widetilde{a}_{n}x+\widetilde{b}_{n}\right)-\Lambda(x)\right|<
\frac {\C_{2}}{\log n}
\end{eqnarray*}
for some absolute constants $0<\C_{1}<\C_{2}$ with normalized
constants $\widetilde{a}_{n}$ and $\widetilde{b}_{n}$ determined by
\begin{eqnarray*}
2\pi \widetilde{b}_{n}^{2}\exp \left( \widetilde{b}^{2}_{n} \right) = n^{2},
\quad
\widetilde{a}_{n}=\widetilde{b}_{n}^{-1}.
\end{eqnarray*}
The following more informative result was established by Nair (1981):
\begin{eqnarray*}
\overline{b}_{n}^{2}\Big( \overline{b}_{n}^{2}\left(
\Phi^{n} \left(\overline{a}_{n}x+\overline{b}_{n}\right) - \Lambda(x) \right) -
\overline{\kappa}(x)\Lambda(x) \Big) \to \left( \overline{\omega}(x)
+\frac {\overline{\kappa}^{2}(x)}{2} \right)\Lambda(x),
\end{eqnarray*}
where the normalized constants $\overline{a}_{n}$ and $\overline{b}_{n}$ are given by
\begin{eqnarray*}
1-\Phi \left(\overline{b}_{n}\right)=n^{-1},
\quad
\overline{a}_{n}=\overline{b}_{n}^{-1},
\end{eqnarray*}
where $\overline{\kappa}(x)$ and $\overline{\omega}(x)$ are defined as
\begin{eqnarray*}
\overline{\kappa}(x)=2^{-1}\left(x^{2}+2x\right)e^{-x}
\quad
\mbox{and}
\quad
\overline{\omega}(x)=
-8^{-1}\left(x^{4}+4x^{3}+8x^{2}+16x\right)e^{-x}.
\end{eqnarray*}

The contents of this note are organized as follows.
Section \ref{sec2} derives some preliminary results related to $\SN(\lambda)$
like Mills inequalities, Mills ratios, and the distributional
tail representation of $\SN(\lambda)$ for $\lambda\ne 0$.
Convergence
rates of the distribution of the maxima of $\SN(\lambda)$ and related proofs
are given in Section \ref{sec3}.
In the sequel we shall assume that the shape parameter $\lambda\ne 0$.

\section{Preliminary results}
\label{sec2}

In this section, some preliminary but important  properties about $\SN(\lambda)$ are derived.
These properties not only imply that $\SN(\lambda)$ belongs to the max-domain of
attraction of the Gumbel extreme value distribution but they also help us to find two pairs of norming constants.

The following Mills inequality and Mills ratio about $\SN(0)$ due to
Mills (1926) are needed in this section, i.e.,
\begin{eqnarray}
\label{eq1.15}
x^{-1}\left( 1+x^{-2} \right)^{-1} \phi(x) < 1-\Phi(x) < x^{-1}
\phi(x)
\end{eqnarray}
for all $x>0$ and
\begin{eqnarray}
\label{eq1.16}
\frac {1-\Phi(x)}{\phi(x)} \sim \frac {1}{x}
\end{eqnarray}
as $x\to \infty$.
For some improved Mills inequalities, see
Mitrinovi\'{c} and Vasi\'{c} (1970, pages 177-180) and references
therein.

First we derive Mills inequalities and Mills ratios of
$\SN(\lambda)$, which are stated as follows.

\begin{proposition}
\label{pro1.1}
Let $F_{\lambda}(x)$ and $f_{\lambda}(x)$ denote the cdf and the pdf of $\SN(\lambda)$, respectively.
For all $x>0$, we have
\begin{itemize}

\item[(i).]
if $\lambda >0$,
\begin{eqnarray}
\label{eq1.1}
x^{-1}\left( 1+x^{-2} \right)^{-1} <
\frac {1-F_{\lambda}(x)}{f_{\lambda}(x)} < x^{-1}\left(
1-\frac {\phi\left( \lambda x \right)}{\lambda x} \right)^{-1},
\end{eqnarray}
which implies
\begin{eqnarray}
\label{eq1.3}
\frac {1-F_{\lambda}(x)}{f_{\lambda}(x)} \sim \frac {1}{x}
\end{eqnarray}
as $x \to \infty$;

\item[(ii).]
if $\lambda <0$,
\begin{eqnarray}
\label{eq1.2}
&&
x^{-1}\left( 1+x^{-2} \right)^{-1}\left(
1-\frac {\lambda^{2}}{1+\lambda^{2}}\left(
1+\frac {1}{\lambda^{2}x^{2}} \right) \right)
\nonumber
\\
&&
<\frac {1-F_{\lambda}(x)}{f_{\lambda}(x)}
\nonumber
\\
&&
<x^{-1}\left( 1-\frac {\lambda^{2}}{1+\lambda^{2}}\left(
1+\frac {1}{\left(1+\lambda^{2} \right)x^{2}} \right)^{-1} \right),
\end{eqnarray}
which implies
\begin{eqnarray}
\label{eq1.4}
\frac {1-F_{\lambda}(x)}{f_{\lambda}(x)} \sim \frac {1}{\left( 1+\lambda^{2} \right)x}
\end{eqnarray}
as $x\to \infty$.

\end{itemize}
\end{proposition}

\noindent
{\bf Proof.}
For $x>0$,
\begin{eqnarray}
\label{eq4.1}
&&
\frac {1}{x^{2}}\int_{x}^{\infty} \Phi(\lambda t)\exp
\left(-\frac {t^{2}}{2}\right) dt
\nonumber
\\
&&
>\int_{x}^{\infty}
\frac {1}{t^{2}}\Phi(\lambda t)\exp \left(
-\frac {t^{2}}{2}\right) dt
\nonumber
\\
&=&
x^{-1}\Phi(\lambda x)\exp \left( -\frac {x^{2}}{2} \right) -
\int_{x}^{\infty} \Phi(\lambda t)\exp\left(-\frac {t^{2}}{2} \right) dt
\nonumber
\\
&&
\qquad\qquad+
\frac {\lambda}{\sqrt{2\pi}}\int_{x}^{\infty}
t^{-1}\exp\left( -\frac {1+\lambda^{2}}{2}t^{2}\right) dt.
\end{eqnarray}
Hence,
\begin{eqnarray}
\label{eq4.2}
&&
\left( 1+\frac {1}{x^{2}} \right) \int_{x}^{\infty} \Phi(\lambda t)
\exp\left( -\frac {t^{2}}{2}  \right)dt
\nonumber
\\
&&
> x^{-1}\Phi(\lambda x)\exp
\left( -\frac {x^{2}}{2} \right) +
\frac {\lambda}{\sqrt{2\pi}}\int_{x}^{\infty} t^{-1}\exp\left(
-\frac {1+\lambda^{2}}{2}t^{2}\right) dt.
\end{eqnarray}

In the case of $\lambda>0$, by \eqref{eq4.2}, we can get
\begin{eqnarray*}
\left( 1+\frac {1}{x^{2}} \right) \int_{x}^{\infty} \Phi(\lambda t)
\exp\left( -\frac {t^{2}}{2}  \right)dt  > x^{-1}\Phi(\lambda x)\exp
\left( -\frac {x^{2}}{2} \right),
\end{eqnarray*}
i.e.,
\begin{eqnarray}
\label{eq4.3}
\int_{x}^{\infty} \Phi(\lambda t)
\exp\left( -\frac {t^{2}}{2}  \right)dt  > x^{-1}\left(
1+\frac {1}{x^{2}} \right)^{-1}\Phi(\lambda x)\exp \left(-\frac {x^{2}}{2} \right).
\end{eqnarray}
By using \eqref{eq4.1} and \eqref{eq1.15}, we have
\begin{eqnarray}
\label{eq4.4}
&&
\int_{x}^{\infty} \Phi(\lambda t)\exp\left(-\frac {t^{2}}{2}
\right) dt
\nonumber
\\
&&<
x^{-1}\Phi(\lambda x)\exp \left( -\frac {x^{2}}{2} \right) +
\frac {\lambda}{\sqrt{2\pi}}\int_{x}^{\infty} t^{-1}\exp\left(
-\frac {1+\lambda^{2}}{2}t^{2}\right) dt
\nonumber
\\
&&<
x^{-1}\Phi(\lambda x)\exp \left( -\frac {x^{2}}{2} \right) \left(
1+ \frac {\frac {\lambda}{\sqrt{2\pi}} \int_{x}^{\infty} \exp\left(
-\frac {1+\lambda^{2}}{2}t^{2}\right) dt}{\Phi(\lambda x)\exp \left(
-\frac {x^{2}}{2}\right)}\right)
\nonumber
\\
&&<
x^{-1}\Phi(\lambda x)\exp \left( -\frac {x^{2}}{2} \right) \left(
1+ \frac {\frac {\lambda}{\sqrt{2\pi}} \int_{x}^{\infty} \exp\left(
-\frac {\lambda^{2}t^{2}}{2}\right) dt}{\Phi(\lambda x)}\right)
\nonumber
\\
&&<
x^{-1}\Phi(\lambda x)\exp \left( -\frac {x^{2}}{2} \right) \left(
1 - \frac {\phi(\lambda x)}{\lambda x}\right)^{-1}.
\end{eqnarray}
Combining \eqref{eq4.3} with \eqref{eq4.4}, we can derive \eqref{eq1.1} and \eqref{eq1.3}.

In the case of $\lambda<0$, by \eqref{eq4.2} and \eqref{eq1.15}, we have
\begin{eqnarray}
\label{eq4.5}
&&
\left( 1+\frac {1}{x^{2}} \right) \int_{x}^{\infty} \Phi(\lambda t)
\exp\left( -\frac {t^{2}}{2}  \right)dt
\nonumber
\\
&&>
x^{-1}\Phi(\lambda x)\exp
\left( -\frac {x^{2}}{2} \right) \left(
1+\frac {\frac {\lambda}{\sqrt{1+\lambda^{2}}}\left(
1-\Phi \left( \sqrt{1+\lambda^{2}}x \right) \right)}{\left( 1-\Phi(-\lambda x)
\right)\exp\left(-\frac {x^{2}}{2}\right)} \right)
\nonumber
\\
&&>
x^{-1}\Phi(\lambda x)\exp \left( -\frac {x^{2}}{2} \right) \left(
1-\frac {\lambda^{2}}{1+\lambda^{2}}\left(
1+\frac {1}{\lambda^{2}x^{2}} \right)\right).
\end{eqnarray}
For $x>0$,
\begin{eqnarray}
\label{eq4.6}
\int_{x}^{\infty} \Phi(\lambda t) \exp\left( -\frac {t^{2}}{2}
\right)dt < x^{-1}\Phi(\lambda x)\exp \left( -\frac {x^{2}}{2}
\right) \left( 1-\frac {\lambda^{2}}{1+\lambda^{2}}\left(
1+\frac {1}{\left( 1+\lambda^{2}\right)x^{2}} \right)^{-1}\right).
\end{eqnarray}
Combining \eqref{eq4.5} with \eqref{eq4.6}, we can derive \eqref{eq1.2} and \eqref{eq1.4}.
\qed

By Proposition \ref{pro1.1}, we can derive  the distributional tail
representation of the skew normal distribution.
This representation is useful to find
optimal normalized constants to establish expansions of the
distribution of the maxima for $\SN(\lambda)$ samples.
Similar expansions for  $\SN(0)$ samples were given in Nair (1981).

\begin{proposition}
\label{pro1.2}
Let $F_{\lambda}(x)$ and $f_{\lambda}(x)$
denote the cdf and the pdf of $\SN(\lambda)$, respectively.
Then,
\begin{eqnarray*}
1-F_{\lambda}(x) = c(x)\exp \left( -\int_{1}^{x} \frac {g(t)}{f(t)}dt \right)
\end{eqnarray*}
for large $x$, where $c(x)$, $g(x)$ and the auxiliary function
$f(x)$ are determined according to the sign of $\lambda$, i.e.,
\begin{itemize}

\item[(i).]
for $\lambda>0$,
\begin{eqnarray*}
c(x) \to \left(\frac {2}{\pi e}\right)^{1/2}
\quad
\mbox{as}
\quad
x \to \infty,
\end{eqnarray*}
\begin{eqnarray}
\label{eq1.7}
f(x)=\frac {1}{x} >0
\quad
\mbox{with}
\quad
f'(x)= -\frac {1}{x^{2}} \to 0
\quad
\mbox{as}
\quad
x\to \infty
\end{eqnarray}
and
\begin{eqnarray}
\label{eq1.8}
g(x)= 1+\frac {1}{x^{2}} \to 1
\quad
\mbox{as}
\quad
x \to \infty.
\end{eqnarray}

\item[(ii).]
for $\lambda<0$,
\begin{eqnarray*}
c(x) \to \frac {\exp \left( -\frac {1+\lambda^{2}}{2} \right)}{(-\lambda) \left(1+\lambda^{2}\right) \pi}
\quad
\mbox{as}
\quad
x \to \infty,
\end{eqnarray*}
\begin{eqnarray}
\label{eq1.10}
f(x)=\frac {1}{(1+\lambda^{2})x} >0
\quad
\mbox{with}
\quad
f'(x) = -\frac {1}{\left(1+\lambda^{2}\right)x^{2}} \to 0
\quad
\mbox{as}
\quad
x \to \infty
\end{eqnarray}
and
\begin{eqnarray}
\label{eq1.11}
g(x) = 1 + \frac {2}{\left(1+\lambda^{2}\right)x^{2}} \to 1
\quad
\mbox{as}
\quad
x\to \infty.
\end{eqnarray}

\end{itemize}
\end{proposition}

Proposition \ref{pro1.2} shows that $F_{\lambda}\in D(\Lambda)$ by
Corollary 1.7 of Resnick (1987).
The norming constants $a_{n}$ and $b_{n}$ can be determined by
\begin{eqnarray}
\label{eq1.12}
1 - F_{\lambda} \left(b_{n}\right)=1/n,
\qquad
a_{n}=f\left(b_{n}\right)
\end{eqnarray}
such that
\begin{eqnarray}
\label{eq1.13}
\lim_{n\to \infty}F^{n}_{\lambda} \left( a_{n}x+b_{n} \right)= \Lambda(x).
\end{eqnarray}

By using Mills ratio of the skew normal distribution and arguments
similar to the case of $\SN(0)$ provided in Leadbetter et al. (1983),
we can choose another pair of normalized constants such that \eqref{eq1.13} holds.

\begin{proposition}
\label{pro1.3}
Let $(X_{n},n\geq 1)$ be a sequence of independent and identically
distributed random variables with common cdf $F_{\lambda}(x)$.
Let $M_{n}=\max(X_{k}, 1\leq k \leq n)$ denote the partial maximum.
Then
\begin{eqnarray}
\label{eq1.14}
\lim_{n\to \infty} \P \left(M_{n} \leq \alpha_{n}x + \beta_{n}\right) = \Lambda(x),
\end{eqnarray}
where the norming constants $\alpha_{n}$ and $\beta_{n}$ are given by
\begin{eqnarray*}
\alpha_{n}=\left( 2\log n \right)^{-\frac {1}{2}},
\quad
\beta_{n}=\left( 2\log n \right)^{\frac {1}{2}}- \frac {\log \log n +\log \pi}{2(2\log n)^{\frac {1}{2}}}
\end{eqnarray*}
for $\lambda>0$, and by
\begin{eqnarray*}
\alpha_{n}=\left(1+\lambda^{2}\right)^{-\frac {1}{2}}(2\log n)^{-\frac {1}{2}},
\quad
\beta_{n}=\left( \frac {2\log n}{1+\lambda^{2}} \right)^{\frac {1}{2}} -
\frac {\log \log n +\log (-2\pi \lambda)}{\left( 1+\lambda^{2} \right)^{\frac {1}{2}}(2\log n)^{\frac {1}{2}} }
\end{eqnarray*}
for $\lambda<0$.
\end{proposition}

\noindent
{\bf Proof.}
First we consider the case of $\lambda>0$.
As $F_{\lambda}$ is continuous, for $x\in R$, there exists
$u_{n}=u_{n}(x)$ such that $n(1-F_{\lambda}(u_{n}))=e^{-x}$.
By using \eqref{eq1.3}, we have
\begin{eqnarray*}
n u_{n}^{-1}\frac {2}{\sqrt{2\pi}}\Phi \left(\lambda u_{n}\right) \exp \left(x - \frac {u_{n}^{2}}{2}\right) \to 1
\end{eqnarray*}
as $n\to \infty$, and so
\begin{eqnarray}
\label{eq4.21}
\log n -\log u_{n} +\frac {1}{2}\left(\log 2-\log\pi\right)
-\frac {u_{n}^{2}}{2} + \log \Phi \left(\lambda u_{n}\right) + x \to 0
\end{eqnarray}
as $n\to \infty$, which implies
\begin{eqnarray*}
\frac {u_{n}^{2}}{2\log n} \to 1
\end{eqnarray*}
as $n\to \infty$.
Taking logarithms, we have
\begin{eqnarray*}
2\log u_{n} - \log 2 -\log \log n \to 0
\end{eqnarray*}
as $n\to \infty$, so
\begin{eqnarray}
\label{eq4.22}
\log u_{n} = \frac {1}{2}\left(\log 2+\log \log n\right) +o(1).
\end{eqnarray}
Noting that $\lim_{n\to \infty} \Phi(\lambda u_{n})=1$ for
$\lambda>0$ and putting \eqref{eq4.22} into \eqref{eq4.21}, we have
\begin{eqnarray*}
u_{n}
&=&
(2\log n)^{\frac {1}{2}} \left( 1-\frac {\log \log n +\log \pi}{2\log n} +\frac {x}{\log n} + o\left(\frac {1}{\log n}
\right)\right)^{\frac {1}{2}}
\\
&=&
(2\log n)^{\frac {1}{2}}\left(  1-\frac {\log \log n +\log
\pi}{4\log n} + \frac {x}{2\log n} +o\left(\frac {1}{\log n} \right) \right)
\\
&=&
\frac {x}{(2\log n)^{\frac {1}{2}}} +(2\log n)^{\frac {1}{2}} -
\frac {\log \log n +\log \pi}{2(2\log n)^{\frac {1}{2}}}+
o\left(\frac {1}{(\log n)^{\frac {1}{2}}} \right)
\\
&=&
\alpha_{n}x +\beta_{n} + o \left(\alpha_{n}\right).
\end{eqnarray*}
So, by Theorem 1.2.3 and Theorem 1.5.1 of Leadbetter et al. (1983), we have
\begin{eqnarray*}
\lim_{n\to \infty} \P \left( M_{n} \leq \alpha_{n}x+\beta_{n} \right) = \Lambda(x)
\end{eqnarray*}
with
\begin{eqnarray*}
\alpha_{n}=(2\log n)^{-\frac {1}{2}}
\end{eqnarray*}
and
\begin{eqnarray*}
\beta_{n}=(2\log n)^{\frac {1}{2}} - \frac {\log \log n +\log \pi}{2(2\log n)^{\frac {1}{2}}}.
\end{eqnarray*}

Similarly, for $\lambda<0$ there exists $v_{n}=v_{n}(x)$ such that
$n(1-F_{\lambda}(v_{n}))=e^{-x}$.
By using \eqref{eq1.4} and \eqref{eq1.16}, we have
\begin{eqnarray*}
nv_{n}^{-2} \left(1+\lambda^{2}\right)^{-1}(-\lambda \pi)^{-1} \exp\left(x - \frac {1+\lambda^{2}}{2}v_{n}^{2}\right) \to 1
\end{eqnarray*}
as $n\to \infty$.
By arguments similar to the case of $\lambda >0$, we can derive
\begin{eqnarray*}
\lim_{n\to \infty} \P \left( M_{n} \leq \alpha_{n}x+\beta_{n} \right) = \Lambda(x)
\end{eqnarray*}
with
\begin{eqnarray*}
\alpha_{n} = \left(1+\lambda^{2}\right)^{-\frac {1}{2}}(2\log n)^{-\frac {1}{2}}
\end{eqnarray*}
and
\begin{eqnarray*}
\beta_{n}=\left(\frac {2\log n}{1+\lambda^{2}}\right)^{\frac {1}{2}} -
\frac {\log \log n +\log (-2\lambda \pi)}{\left(1+\lambda^{2}\right)^{\frac {1}{2}}(2\log n)^{\frac {1}{2}}}.
\end{eqnarray*}
The proof is complete.
\qed

\section{Convergence rates of extremes}
\label{sec3}

In this section, we establish two different convergence rates of the distribution of $M_{n}$:
one for the norming constants $\alpha_{n}$ and $\beta_{n}$ given by
Proposition \ref{pro1.3}, and the other for the norming constants $a_{n}$ and
$b_{n}$ determined by \eqref{eq1.12}.

\begin{theorem}
\label{thm2.1}
For norming constants $\alpha_{n}$ and $\beta_{n}$ given by
Proposition \ref{pro1.3}, we have
\begin{itemize}

\item[(i).]
For $\lambda>0$,
\begin{eqnarray}
\label{eq2.1}
F_{\lambda}^{n} \left( \alpha_{n}x +\beta_{n} \right) - \Lambda(x) \sim
\frac {\Lambda(x)e^{-x}}{16}\frac {(\log \log n)^{2}}{\log n}
\end{eqnarray}
as $n\to \infty$;

\item[(ii).]
For $\lambda<0$,
\begin{eqnarray}
\label{eq2.2}
F_{\lambda}^{n} \left( \alpha_{n}x +\beta_{n} \right) - \Lambda(x) \sim
\frac {\Lambda(x)e^{-x}}{4}\frac {(\log \log n)^{2}}{\log n}
\end{eqnarray}
as $n\to \infty$.

\end{itemize}
\end{theorem}

\noindent
{\bf Proof.}
Set $u_{n}=\alpha_{n}x+\beta_{n}$ and
$\tau_{n}=n(1-F_{\lambda}(u_{n}))$, where $\alpha_{n}$ and
$\beta_{n}$ are given by Proposition \ref{pro1.3}.

For $\lambda>0$,
\begin{eqnarray*}
u_{n}= (2\log n)^{-\frac {1}{2}}x + (2\log n)^{\frac {1}{2}}
-2^{-1}(2\log n)^{-\frac {1}{2}}(\log \log n+\log \pi)
\end{eqnarray*}
implies
\begin{eqnarray*}
u_{n}^{2}=2\left( \log n + x - \frac {\log \log n +\log \pi}{2} +
\frac {(\log \log n)^{2}}{16\log n}(1+o(1)) \right),
\end{eqnarray*}
\begin{eqnarray*}
u_{n}^{-1}=(2\log n)^{-\frac {1}{2}}\left( 1+O\left( \frac {\log \log n}{\log n} \right) \right)
\end{eqnarray*}
and
\begin{eqnarray*}
O\left(u_{n}^{-2}\right)=O\left( \frac {1}{\log n} \right).
\end{eqnarray*}
Noting that  $\log \Phi(\lambda u_{n})=o\left( (\log \log n)^{2}/\log n \right)$ for large $n$
and rewriting \eqref{eq1.1} as
\begin{eqnarray*}
\frac {1}{x}\left( 1 - \frac {1}{1+x^{2}} \right)<
\frac {1-F_{\lambda}(x)}{f_{\lambda}(x)} < \frac {1}{x}\left(
1+\frac {\phi(\lambda x)}{\lambda x - \phi(\lambda x)} \right),
\end{eqnarray*}
we have
\begin{eqnarray*}
\tau_{n}
&=&
n \left( 1 - F_{\lambda} \left( u_{n} \right) \right)
\\
&=&
\frac {2n}{\sqrt{2\pi}u_{n}}\exp \left( -\frac {u_{n}^{2}}{2} +
\log \Phi \left(\lambda u_{n}\right) \right) \left( 1 + O \left(u_{n}^{-2}\right) \right)
\\
&=&
e^{-x}\left( 1 - \frac {(\log \log n)^{2}}{16\log n}(1+o(1)) \right).
\end{eqnarray*}
Obviously, for $\tau(x)=e^{-x}$,
\begin{eqnarray*}
\tau(x) - \tau_{n}(x)=e^{-x}\frac {(\log \log n)^{2}}{16\log n} \left(1+o(1)\right) \sim e^{-x}\frac {(\log \log n)^{2}}{16\log n}
\end{eqnarray*}
for large $n$.
So, by using Theorem 2.4.2 of Leadbetter et al. (1983), we can get \eqref{eq2.1}.

In the case of $\lambda<0$, using \eqref{eq1.2} and the following fact
\begin{eqnarray*}
1-\Phi(x)=\frac {\phi(x)}{x} \left[ 1 + O \left( x^{-2} \right) \right],
\end{eqnarray*}
we can get \eqref{eq2.2} by arguments similar to the case of $\lambda>0$.
\qed

Theorem \ref{thm2.2} shows that the convergence rates of
the distribution of the normalized maxima are different even though both
\eqref{eq1.13} and \eqref{eq1.14} hold, which implies
$\alpha_{n}/a_{n}\to 1$ and $(\beta_{n}-b_{n})/a_{n}\to 0$  by
Khintchine Theorem (cf. Leadbetter et al.  (1983), Resnick (1987)).
The auxiliary function $f(x)$ of the
distributional tail representation of $F_{\lambda}(x)$ plays an
important role on the convergence rates for distributions
belonging to the domain of attraction of the Gumbel extreme value
distribution.
For related work, see Peng et al. (2010) and Liao and Peng (2012):
Peng et al. (2010) derived convergence rates of the distribution of maxima
for random sequences obeying the general error distribution;
Liao and Peng (2012) derived convergence rates of the distribution of maxima
for random sequences obeying the lognormal distribution.
Another interesting application of the auxiliary function $f(x)$ is to characterize $F\in D(\Lambda)$.
For more details, see Goldie and Resnick (1988).

\begin{theorem}
\label{thm2.2}
For norming constants $a_{n}$ and $b_{n}$ given by \eqref{eq1.12}, we have
\begin{eqnarray}
\label{eq2.3}
b_{n}^{2}\Bigg[ b_{n}^{2}\bigg(  F_{\lambda}^{n}\left( a_{n}x +b_{n} \right) - \Lambda(x) \bigg) - \kappa(x)\Lambda(x) \Bigg] \to \left(
\omega(x) + \frac {\kappa^{2}(x)}{2}\right)\Lambda(x)
\end{eqnarray}
as $n\to \infty$, where both $\kappa(x)$ and $\omega(x)$ may
depend on $\lambda$ according to the sign of $\lambda$ as:
\begin{itemize}

\item[(i).]
for $\lambda>0$,
\begin{eqnarray*}
\kappa(x)=2^{-1} \left(x^{2}+2x\right)e^{-x},
\quad
\omega(x)=-8^{-1} \left(x^{4}+4x^{3}+8x^{2}+16x\right)e^{-x};
\end{eqnarray*}

\item[(ii).]
for $\lambda<0$,
\begin{eqnarray*}
\kappa(x)=2^{-1} \left(1+\lambda^{2}\right)^{-1} \left(x^{2}+4x\right)e^{-x}
\end{eqnarray*}
and
\begin{eqnarray*}
\omega(x)=-8^{-1}\lambda^{-2} \left(1+\lambda^{2}\right)^{-2}
\Big[ \lambda^{2}x^{4}+8\lambda^{2}x^{3}+ 24\lambda^{2}x^{2}+16 \left(1+3\lambda^{2}\right)x\Big] e^{-x}.
\end{eqnarray*}

\end{itemize}
\end{theorem}

To prove Theorem \ref{thm2.2}, we need two lemmas.
The first lemma is about a decomposition of the distributional tail representation
of the skew normal distribution.

\begin{lemma}
\label{lem1}
Let $F_{\lambda}(x)$ denote the cdf of the skew normal distribution.
For large $x$, we have
\begin{itemize}

\item[(i).]
for $\lambda >0$,
\begin{eqnarray}
\label{eq4.7}
1-F_{\lambda}(x)=\left(\frac {2}{\pi e}\right)^{\frac {1}{2}}
\Big[ 1-x^{-2}+3x^{-4}+O \left(x^{-6}\right) \Big]
\Phi(\lambda x) \exp \left(
-\int_{1}^{x} \frac {g(t)}{f(t)} dt \right)
\end{eqnarray}
with $f(t)$ and $g(t)$ given by \eqref{eq1.7} and \eqref{eq1.8}, respectively;

\item[(ii).]
for $\lambda <0$,
\begin{eqnarray}
\label{eq4.8}
1-F_{\lambda}(x)
&=&
\frac {\exp \left( -\frac {1+\lambda^{2}}{2} \right)}{(-\lambda)\pi \left(1+\lambda^{2}\right)} \Big[
1 - \frac {1+3\lambda^{2}}{\lambda^{2} \left(1+\lambda^{2}\right)}x^{-2}
\nonumber
\\
&&
\qquad\qquad
+\frac {15\lambda^{4}+10\lambda^{2}+3}{\lambda^{4} \left(1+\lambda^{2}\right)^{2}}x^{-4}
+O\left(x^{-6}\right) \Big]
\exp \left( -\int_{1}^{x} \frac {g(t)}{f(t)} dt \right)
\end{eqnarray}
with $f(t)$ and $g(t)$ given by \eqref{eq1.10} and \eqref{eq1.11}, respectively.
\end{itemize}
\end{lemma}

\noindent
{\bf Proof.}
By integration by parts, we have
\begin{eqnarray}
\label{eq4.9}
1-F_{\lambda}(x)
&=&
\frac {f_{\lambda}(x)}{x} \Bigg[  1 - x^{-2} +
3x^{-4} - 15x^{-6} + \frac {\lambda}{1+\lambda^{2}}
\frac {\phi(\lambda x)}{\Phi(\lambda x)}x^{-1} -
\frac {\lambda \left(3+\lambda^{2}\right)}{\left( 1 + \lambda^{2} \right)^{2}}
\frac {\phi(\lambda x)}{\Phi(\lambda x)}x^{-3}
\nonumber
\\
&&
\qquad\qquad\qquad +
\frac {\lambda}{1+\lambda^{2}}\left( 3+\frac {4 \left(3+\lambda^{2}\right)}{\left(1+\lambda^{2}\right)^{2}} \right)
\frac {\phi(\lambda x)}{\Phi(\lambda x)}x^{-5}
\nonumber
\\
&&
\qquad\qquad\qquad -
\frac {\lambda}{1+\lambda^{2}}\left(15+\frac {6}{1+\lambda^{2}}
\left( 3+\frac {4 \left( 3+\lambda^{2}\right)}
{\left(1+\lambda^{2}\right)^{2}}\right)\right)
\frac {\phi(\lambda x)}{\Phi(\lambda x)}x^{-7} \Bigg]
\nonumber
\\
&&
+\frac {210}{\sqrt{2\pi}} \int_{x}^{\infty} t^{-8}\Phi(\lambda t)
\exp\left( -\frac {t^{2}}{2} \right) dt
\nonumber
\\
&&
+\frac {8\lambda}{\pi \left( 1+\lambda^{2} \right)}
\left( 15+\frac {18}{1+\lambda^{2}} \right)
\int_{x}^{\infty} t^{-9} \exp \left( -\frac {1+\lambda^{2}}{2}t^{2} \right) dt
\nonumber
\\
&&
+\frac {192 \lambda \left(3+\lambda^{2}\right)}{\pi \left(1+\lambda^{2}\right)^4}
\int_{x}^{\infty} t^{-9} \exp \left( -\frac {1+\lambda^{2}}{2}t^{2} \right) dt.
\end{eqnarray}
It is easy to check that both
\begin{eqnarray}
\label{addpeng2}
0< \int_{x}^{\infty} t^{-8}\Phi(\lambda t)\exp\left( -\frac {t^{2}}{2} \right) dt
< x^{-9}\Phi(\lambda x) \exp\left( -\frac {x^{2}}{2}\right) +
\frac {\lambda}{\sqrt{2\pi}}\int_{x}^{\infty} t^{-9} \exp \left( -\frac {1+\lambda^{2}}{2}t^{2} \right) dt
\end{eqnarray}
and
\begin{eqnarray}
\label{addpeng3}
0< \int_{x}^{\infty} t^{-9} \exp \left( -\frac {1+\lambda^{2}}{2}t^{2} \right) dt <
\frac {x^{-10}}{1+\lambda^{2}}\exp\left( -\frac {1+\lambda^{2}}{2}x^{2}\right)
\end{eqnarray}
hold for $x>0$.

For $\lambda>0$, it is obvious that $x^{5}\phi(\lambda x)\to 0$ and
$\Phi(\lambda x) \to 1$ as $x \to \infty$.
So, by \eqref{eq4.9}, \eqref{addpeng2} and \eqref{addpeng3}, we can get
\begin{eqnarray*}
1-F_{\lambda}(x)
&=&
\frac {f_{\lambda}(x)}{x} \left(
1-x^{-2}+3x^{-4}+O \left(x^{-6}\right) \right)
\nonumber
\\
&=&
\left(\frac {2}{\pi e}\right)^{\frac {1}{2}} \Big(
1-x^{-2}+3x^{-4}+O \left(x^{-6}\right) \Big) \Phi(\lambda x)
\exp \left( -\int_{1}^{x} t\left( 1+\frac {1}{t^{2}}\right) dt \right),
\end{eqnarray*}
which is \eqref{eq4.7}.

In the case of $\lambda<0$, first notice that
\begin{eqnarray}
\label{eq1.17}
1 - \Phi(x) = \frac {\phi(x)}{x} \left[ 1-x^{-2} +3x^{-4} +O\left( x^{-6}\right) \right]
\end{eqnarray}
for large $x$, cf. Castro (1987).
So, by \eqref{eq4.9} and \eqref{eq1.17}, we have
\begin{eqnarray*}
1-F_{\lambda}(x)
&=&
\frac {2\phi(x)}{x} \Bigg[ \Phi(\lambda x)\left(
1-x^{-2}+3x^{-4} + O \left(x^{-6}\right) \right) +
\frac {\lambda}{1+\lambda^{2}}\frac {\phi(\lambda x)}{x} -
\frac {\lambda \left(3+\lambda^{2}\right)}{\left(1+\lambda^{2}\right)^{2}}
\frac {\phi(\lambda x)}{x^{3}}
\\
&&
\qquad\qquad\qquad+
\frac {\lambda}{1+\lambda^{2}}\left(
3 + \frac {4 \left(3+\lambda^{2}\right)}{\left(1+\lambda^{2}\right)^{2}} \right)
\frac {\phi(\lambda x)}{x^{5}} + O\left( \frac {\phi(\lambda x)}{x^{7}} \right) \Bigg]
\\
&=&
\frac {\exp \left( -\frac {1+\lambda^{2}}{2}x^{2} \right)}
{(-\lambda)\pi \left(1+\lambda^{2}\right)x^{2}}
\left( 1 - \frac {1+3\lambda^{2}}{\lambda^{2} \left(1+\lambda^{2}\right)} x^{-2} +
\frac {15\lambda^{4}+10\lambda^{2}+3}{\lambda^{4} \left(1+\lambda^{2}\right)^{2}}x^{-4} + O \left(x^{-6}\right) \right)
\nonumber
\\
&=&
\frac {\exp \left( -\frac {1+\lambda^{2}}{2} \right)}{(-\lambda)\pi \left(1+\lambda^{2}\right)}
\Big( 1 - \frac {1+3\lambda^{2}}{\lambda^{2} \left( 1+\lambda^{2} \right)}x^{-2} +
\frac {15\lambda^{4}+10\lambda^{2}+3}{\lambda^{4} \left(1+\lambda^{2}\right)^{2}}x^{-4}
\\
&&
\qquad\qquad\qquad
+ O \left(x^{-6}\right) \Big)
\exp \left(
-\int_{1}^{x} \left(1+\lambda^{2}\right)t\left(
1+\frac {2}{\left(1+\lambda^{2}\right)t^{2}} \right) dt \right),
\end{eqnarray*}
which is \eqref{eq4.8}.
The proof is complete.
\qed

To prove \eqref{eq2.3}, we need the following auxiliary result.

\begin{lemma}
\label{lem2}
Let $H_{\lambda}(b_{n};x)=F_{\lambda}(a_{n}x+b_{n})$ and
$h_{\lambda}(b_{n};x)=n\log H_{\lambda}(b_{n};x) + e^{-x}$ with the
normalized constants $a_{n}$ and $b_{n}$ given by \eqref{eq1.12}.
Then,
\begin{eqnarray}
\label{eq4.10}
\lim_{n \to \infty} b_{n}^{2}\Big( b_{n}^{2}h_{\lambda} \left(b_{n}; x \right) - \kappa(x) \Big) =\omega(x),
\end{eqnarray}
where $\kappa(x)$ and $\omega(x)$ are those given by Theorem \ref{thm2.2}.
\end{lemma}

\noindent
{\bf Proof.}
First, we consider the case of $\lambda>0$.
Obviously, $b_{n} \to \infty$ if and only if $n\to \infty$ since
$1-F_{\lambda}(b_{n})=n^{-1}$.
The following two facts hold by Proposition \ref{pro1.1}:
\begin{eqnarray}
\label{eq4.11}
\lim_{n\to \infty} \frac {b_{n}\left( 1-F_{\lambda} \left( b_{n} + b_{n}^{-1}x \right) \right)}
{f_{\lambda} \left(b_{n}\right)} = e^{-x}
\end{eqnarray}
and
\begin{eqnarray}
\label{eq4.12}
\lim_{n \to \infty}
\frac {1-F_{\lambda} \left( b_{n}+b_{n}^{-1}x \right)}{b_{n}^{-4}} = 0.
\end{eqnarray}
For simplicity, let
\begin{eqnarray*}
A_{\lambda} \left( b_{n} \right) = \frac {\Phi \left(\lambda b_{n} \right)
\left( 1 - b_{n}^{-2}+3b_{n}^{-4} + O \left(b_{n}^{-6}\right)\right)}
{\Phi\left(\lambda \left(b_{n}+b_{n}^{-1}x\right)\right)
\left(1 - \left(b_{n}+b_{n}^{-1}x\right)^{-2} + 3 \left( b_{n}+b_{n}^{-1}x \right)^{-4} +
O\left(\left(b_{n}+b_{n}^{-1}x\right)^{-6}\right)\right)}.
\end{eqnarray*}
Then $\lim_{n\to \infty}A_{\lambda}(b_{n}) = 1$ and
\begin{eqnarray*}
A_{\lambda} \left(b_{n}\right) - 1
&=&
\left( 1+o(1) \right) \Big[  \Phi \left( \lambda b_{n} \right) - \Phi \left( \lambda \left( b_{n} +b_{n}^{-1}x \right) \right) +
b_{n}^{-2}\left( \Phi \left( \lambda \left( b_{n}+b_{n}^{-1}x \right) \right) - \Phi \left( \lambda b_{n} \right) \right)
\\
&&
-2xb_{n}^{-4}\Phi \left( \lambda \left( b_{n}+b_{n}^{-1}x \right) \right) +
3b_{n}^{-4}\left( \Phi \left( \lambda b_{n} \right) - \Phi \left( \lambda \left( b_{n}+b_{n}^{-1}x \right) \right)
\left( 1+b_{n}^{-2}x \right)^{-4} \right)
\\
&&
+ O \left( b_{n}^{-6} \right) \Big].
\end{eqnarray*}
So, we can check that both
\begin{eqnarray}
\label{eq4.13}
\lim_{n \to \infty} \frac {A_{\lambda} \left(b_{n}\right) - 1}{b_{n}^{-2}} =0
\end{eqnarray}
and
\begin{eqnarray}
\label{eq4.14}
\lim_{n\to \infty} \frac {A_{\lambda} \left(b_{n}\right) - 1}{b_{n}^{-4}} = -2x
\end{eqnarray}
hold.
By \eqref{eq4.7}, we have
\begin{eqnarray}
\label{eq4.15}
\frac {1 - F_{\lambda} \left(b_{n}\right)}{1 - F_{\lambda} \left(b_{n}+b_{n}^{-1}x\right)}e^{-x}
&=&
A_{\lambda} \left(b_{n}\right)
\exp \left( \int_{0}^{x}\left( b_{n}^{-2}y
+\frac {1}{b_{n}^{2} \left(1+b_{n}^{-2}y\right)} \right)dy \right)
\nonumber
\\
&=&
A_{\lambda} \left(b_{n}\right)
\Big[ 1+ \int_{0}^{x}\left( b_{n}^{-2}y
+\frac {1}{b_{n}^{2} \left(1+b_{n}^{-2}y\right)} \right)dy
\nonumber
\\
&&
+\frac {1}{2}\left( \int_{0}^{x}\left( b_{n}^{-2}y
+\frac {1}{b_{n}^{2} \left(1+b_{n}^{-2}y\right)} \right)dy \right)^{2} \left(1+o(1)\right) \Big].
\end{eqnarray}
Combining \eqref{eq1.3}, \eqref{eq4.11}, \eqref{eq4.12},
\eqref{eq4.13}, \eqref{eq4.14} and \eqref{eq4.15} together, we have
\begin{eqnarray}
\label{addpeng1}
\lim_{n\to \infty} b_{n}^{2}h_{\lambda} \left(b_{n}; x\right)
&=&
\lim_{n\to \infty} \frac {\log H_{\lambda} \left(b_{n}; x\right) + n^{-1}e^{-x}}{n^{-1}b_{n}^{-2}}
\nonumber
\\
&=&
\lim_{n\to \infty} \frac {\log F_{\lambda} \left(b_{n}+b_{n}^{-1}(x)\right) +
\left[ 1 - F_{\lambda} \left(b_{n}\right) \right]e^{-x}}
{f_{\lambda} \left(b_{n}\right)b_{n}^{-3}}
\nonumber
\\
&=&
\lim_{n\to \infty} \frac {-\left[ 1 - F_{\lambda} \left( b_{n}+b_{n}^{-1}x \right) \right] -
\frac {1}{2} \left[ 1 - F_{\lambda} \left( b_{n}+b_{n}^{-1}x \right) \right]^{2} \left(1+o(1)\right)}
{f_{\lambda} \left( b_{n} \right)b_{n}^{-3}}
\nonumber
\\
&&
+\lim_{n\to \infty} \frac {\left[ 1 - F_{\lambda} \left( b_{n} \right) \right] e^{-x}}{f_{\lambda} \left( b_{n} \right)b_{n}^{-3}}
\nonumber
\\
&=&
\lim_{n\to \infty}
\frac {b_{n} \left[ 1 - F_{\lambda} \left( b_{n}+b_{n}^{-1}x \right) \right]}{f_{\lambda} \left(b_{n}\right)}
\frac {\frac {1 - F_{\lambda} \left(b_{n}\right)}{1 - F_{\lambda} \left(b_{n}+b_{n}^{-1}x \right)}e^{-x} - 1}{b_{n}^{-2}}
\nonumber
\\
&=&
e^{-x} \lim_{n\to \infty} \frac {A_{\lambda} \left(b_{n}\right) - 1 + A_{\lambda} \left(b_{n}\right)
\int_{0}^{x}\left( b_{n}^{-2}y +\frac {1}{b_{n}^{2} \left( 1 + b_{n}^{-2}y \right)} \right)dy
\left( 1 + o(1) \right)}{b_{n}^{-2}}
\nonumber
\\
&=&
e^{-x} \lim_{n\to \infty} \int_{0}^{x} \left( y + \frac {1}{1+b_{n}^{-2}y} \right)dy
\nonumber
\\
&=&
2^{-1} \left(x^{2}+2x\right)e^{-x} = \kappa(x),
\end{eqnarray}
where the last step follows by the dominated convergence
theorem.
It remains  to prove \eqref{eq4.10} for $\lambda>0$.
By arguments similar to those of \eqref{addpeng1}, we have
\begin{eqnarray*}
&&
\lim_{n\to \infty} b_{n}^{2}\Big[ b_{n}^{2}h_{\lambda} \left( b_{n}; x \right) - \kappa(x) \Big]
\\
&=&
\lim_{n\to \infty} \frac {\log H_{\lambda} \left(b_{n}; x\right) +  n^{-1}e^{-x}-b_{n}^{-2}n^{-1}\kappa(x)}
{n^{-1}b_{n}^{-4}}
\\
&=&
\lim_{n\to \infty} \frac {\log F_{\lambda} \left(b_{n}+b_{n}^{-1}x\right) +  \left[ 1 - F_{\lambda} \left(b_{n}\right) \right] e^{-x}
\left( 1 - b_{n}^{-2}\frac {x^{2}+2x}{2} \right)}{f_{\lambda} \left( b_{n} \right)b_{n}^{-5}}
\\
&=&
\lim_{n\to \infty}
\frac {b_{n} \left[ 1 - F_{\lambda} \left( b_{n}+b_{n}^{-1}x \right) \right]}{f_{\lambda} \left(b_{n}\right)}
\frac {\frac {1-F_{\lambda} \left( b_{n} \right)}{1 - F_{\lambda} \left( b_{n}+b_{n}^{-1}x \right)} e^{-x}
\left( 1 - b_{n}^{-2}\frac {x^{2}+2x}{2} \right) - 1}{b_{n}^{-4}}
\\
&=&
e^{-x}\lim_{n\to \infty} \Bigg[
A_{\lambda} \left(b_{n}\right) b_{n}^{2}\left(\int_{0}^{x}\left( y + \frac {1}{1+b_{n}^{-2}y}\right)dy -
\frac {x^{2}+2x}{2} \right)
\\
&&
\qquad\qquad -
A_{\lambda} \left(b_{n}\right) \frac {x^{2}+2x}{2}\int_{0}^{x}
\left( y + \frac {1}{1+b_{n}^{-2}y}\right)dy
\\
&&
\qquad\qquad +
A_{\lambda} \left( b_{n} \right)
\left(\int_{0}^{x}\left( y + \frac {1}{1+b_{n}^{-2}y}\right)dy \right)^{2}
\left( \frac {1}{2} - \frac {x^{2}+2x}{4b_{n}^{2}}\right) \left(1+o(1)\right)
\\
&&
\qquad\qquad +
\frac {A_{\lambda} \left( b_{n} \right) - 1}{b_{n}^{-4}} \Bigg]
\\
&=&
\left( -\frac {x^{2}}{2} -
\frac {1}{2}\left( \frac {x^{2}+2x}{2} \right)^{2} - 2x \right)e^{-x}
\\
&=&
-8^{-1} \left( x^{4}+4x^{3}+8x^{2}+16x \right) e^{-x} =\omega(x).
\end{eqnarray*}

Second, we consider the case of $\lambda<0 $.
By \eqref{eq1.4}, \eqref{eq1.10}, and arguments similar to the case of $\lambda>0$,  we can see that
\begin{eqnarray}
\label{eq4.16}
\lim_{n\to \infty}
\frac {b_{n}\left[ 1-F_{\lambda}\left(b_{n}+ \left( 1+\lambda^{2} \right)^{-1}b_{n}^{-1}x\right) \right]}
{\left( 1 + \lambda^{2} \right)^{-1}f_{\lambda} \left(b_{n}\right)} = e^{-x}
\end{eqnarray}
and
\begin{eqnarray}
\label{eq4.17}
\lim_{n\to \infty}
\frac {1-F_{\lambda}\left(b_{n} + \left(1+\lambda^{2}\right)^{-1}b_{n}^{-1}x\right)}{b_{n}^{-4}} = 0
\end{eqnarray}
hold.
Setting
\begin{eqnarray*}
B_{\lambda} \left(b_{n}\right) = \frac {1 - \frac {1+3\lambda^{2}}{\lambda^{2} \left(1+\lambda^{2}\right)}b_{n}^{-2} +
\frac {15\lambda^{4}+10\lambda^{2}+3}{\lambda^{4} \left(1+\lambda^{2}\right)^{2}}b_{n}^{-4}+O\left(b_{n}^{-6}\right)}
{1 - \frac {1+3\lambda^{2}}{\lambda^{2} \left(1+\lambda^{2}\right)} \left[b_{n} + \left(1+\lambda^{2}\right)^{-1}b_{n}^{-1}x\right]^{-2}
+\frac {15\lambda^{4}+10\lambda^{2}+3}{\lambda^{4}\left(1+\lambda^{2}\right)^{2}}
\left[ b_{n} + \left(1+\lambda^{2}\right)^{-1}b_{n}^{-1}x\right]^{-4}+O \left(b_{n}^{-6}\right)},
\end{eqnarray*}
we have $\lim_{n \to \infty}B_{\lambda}(b_{n})=1$ and
\begin{eqnarray*}
B_{\lambda} \left(b_{n}\right) - 1 = \left(1+o(1)\right)
\left[ -\frac {2\left(1+3\lambda^{2}\right)x}{\lambda^{2}\left(1+\lambda^{2}\right)^{2}}b_{n}^{-4} + O \left(b_{n}^{-6}\right) \right],
\end{eqnarray*}
which implies
\begin{eqnarray}
\label{eq4.18}
\lim_{n\to \infty} \frac {B_{\lambda} \left(b_{n}\right) - 1}{b_{n}^{-2}} = 0
\end{eqnarray}
and
\begin{eqnarray}
\label{eq4.19}
\lim_{n\to \infty} \frac {B_{\lambda} \left(b_{n}\right) - 1}{b_{n}^{-4}}=
-\frac {2 \left(1+3\lambda^{2}\right)x}{\lambda^{2} \left( 1+\lambda^{2} \right)^{2}}.
\end{eqnarray}
By \eqref{eq4.8}, we have
\begin{eqnarray}
\label{eq4.20}
&&
\frac {1 - F_{\lambda} \left(b_{n}\right)}{1 - F_{\lambda} \left(b_{n} + \left(1+\lambda^{2}\right)^{-1}b_{n}^{-1}x\right)} e^{-x}
\nonumber
\\
&=&
B_{\lambda} \left(b_{n}\right)
\exp \left( \int_{0}^{x} \left(
\frac {y}{\left(1+\lambda^{2}\right)b_{n}^{2}} +
\frac {2}{1+\lambda^{2}}
\frac {1}{b_{n}^{2} \left[ 1 + \left(1+\lambda^{2}\right)^{-1}b_{n}^{-2}y \right]} \right)dy \right)
\nonumber
\\
&=&
B_{\lambda} \left(b_{n}\right)
\Bigg[  1+ \int_{0}^{x} \left(
\frac {y}{\left(1+\lambda^{2}\right)b_{n}^{2}} +
\frac {2}{1+\lambda^{2}}
\frac {1}{b_{n}^{2} \left[ 1 + \left(1+\lambda^{2}\right)^{-1}b_{n}^{-2}y \right]} \right)dy
\nonumber
\\
&&
+ \frac {1}{2}\left( \int_{0}^{x} \left(
\frac {y}{\left( 1+\lambda^{2} \right)b_{n}^{2}} +
\frac {2}{1+\lambda^{2}}
\frac {1}{b_{n}^{2}\left[ 1 + \left(1+\lambda^{2}\right)^{-1}b_{n}^{-2}y\right]} \right)dy  \right)^{2} \left(1+o(1)\right) \Bigg].
\end{eqnarray}
Combining \eqref{eq1.4}, \eqref{eq4.16}, \eqref{eq4.17},
\eqref{eq4.18}, \eqref{eq4.19} and \eqref{eq4.20} together, we have
\begin{eqnarray*}
\lim_{n\to \infty} b_{n}^{2}h_{\lambda} \left(b_{n}; x\right)
&=&
\lim_{n\to \infty}  \frac {b_{n}\left[ 1 - F_{\lambda}\left(b_{n} + \left(1+\lambda^{2}\right)^{-1}b_{n}^{-1}x\right) \right]}
{\left( 1  + \lambda^{2} \right)^{-1}f_{\lambda} \left(b_{n}\right)}
\frac {\frac {1-F_{\lambda} \left(b_{n}\right)}{1 - F_{\lambda} \left( b_{n} + \left(1+\lambda^{2}\right)^{-1}b_{n}^{-1}x\right)}
e^{-x} -1}{b_{n}^{-2}}
\\
&=&
e^{-x} \lim_{n\to \infty} \frac {B_{\lambda} \left(b_{n}\right) - 1 + B_{\lambda} \left(b_{n}\right)
\int_{0}^{x} \left( \frac {y}{\left(1+\lambda^{2}\right) b_{n}^{2}} +
\frac {{2}/\left( 1 + \lambda^{2} \right)}
{b_{n}^{2} \left[ 1 + \left(1+\lambda^{2}\right)^{-1}b_{n}^{-2}y\right]} \right)dy}{b_{n}^{-2}}
\\
&=&
2^{-1} \left(1+\lambda^{2}\right)^{-1} \left( x^{2}+4x \right)e^{-x} = \kappa(x)
\end{eqnarray*}
and
\begin{eqnarray*}
&&
\lim_{n\to \infty} b_{n}^{2}\Big[ b_{n}^{2}h_{\lambda} \left( b_{n}; x \right) - \kappa(x) \Big]
\\
&=&
\lim_{n\to \infty} \frac {b_{n}\left[  1 - F_{\lambda}\left(b_{n} + \left(1+\lambda^{2}\right)^{-1}b_{n}^{-1}x\right) \right]}
{\left(1+\lambda^{2}\right)^{-1}f_{\lambda} \left(b_{n}\right)}
\frac {\frac {1 - F_{\lambda} \left(b_{n}\right)}{1 - F_{\lambda} \left(b_{n} + \left(1+\lambda^{2}\right)^{-1}b_{n}^{-1}x \right)}
e^{-x}\left[ 1 - b_{n}^{-2}\frac {x^{2}+4x}{2 \left(1+\lambda^{2}\right)} \right] - 1}{b_{n}^{-4}}
\\
&=&
e^{-x} \lim_{n\to \infty} \Bigg[
\frac {B_{\lambda} \left(b_{n}\right) - 1}{b_{n}^{-4}} +
B_{\lambda} \left(b_{n}\right) b_{n}^{2} \int_{0}^{x} \left( \frac {y}{1+\lambda^{2}} +
\frac {2}{1+\lambda^{2}}\frac {1}{1 + \left(1+\lambda^{2}\right)^{-1}b_{n}^{-2}y} \right)dy
\\
&&
-B_{\lambda} \left(b_{n}\right) b_{n}^{2} \frac {x^{2}+4x}{2\left(1+\lambda^{2}\right)}
\\
&&
-B_{\lambda} \left(b_{n}\right)
\frac {x^{2}+4x}{2 \left(1+\lambda^{2}\right)}
\int_{0}^{x} \left( \frac {y}{1+\lambda^{2}} +
\frac {2}{1+\lambda^{2}}\frac {1}{1 + \left(1+\lambda^{2}\right)^{-1} b_{n}^{-2}y}
\right)dy
\\
&&
+ \frac {1}{2}B_{\lambda} \left(b_{n}\right)
\left(\int_{0}^{x} \left( \frac {y}{1+\lambda^{2}}
+\frac {2}{1+\lambda^{2}}\frac {1}{1 + \left(1+\lambda^{2}\right)^{-1}b_{n}^{-2}y} \right)dy\right)^{2}
\left( 1 + o(1) \right)
\\
&&
- \frac {1}{2} b_n^{-2} B_{\lambda} \left(b_{n}\right)
\left(\int_{0}^{x} \left( \frac {y}{1+\lambda^{2}}
+\frac {2}{1+\lambda^{2}}\frac {1}{1 + \left(1+\lambda^{2}\right)^{-1}b_{n}^{-2}y} \right)dy\right)^{2}
\frac {x^{2}+4x}{2 \left(1+\lambda^{2}\right)} \left( 1+o(1) \right) \Bigg]
\\
&=&
e^{-x}\left[ -\frac {2 \left(1+3\lambda^{2}\right)}{\lambda^{2} \left(1+\lambda^{2}\right)^{2}} x -
\frac {x^{2}}{\left(1+\lambda^{2}\right)^{2}} - \frac {1}{2} \left( \frac {x^{2}+4x}{2 \left(1+\lambda^{2}\right)} \right)^{2} \right]
\\
&=&
-8^{-1}\lambda^{-2} \left(1+\lambda^{2}\right)^{-2}
\left[ \lambda^{2}x^{4}+8\lambda^{2}x^{3}+24\lambda^{2}x^{2}+16 \left(1+3\lambda^{2}\right) x\right] e^{-x} = \omega(x).
\end{eqnarray*}
The claimed result follows for $\lambda<0$.
The proof is complete.
\qed

\noindent
{\bf Proof of Theorem \ref{thm2.2}.}
Obviously, by Lemma \ref{lem2}, we have $h_{\lambda}(b_{n};x)\to 0$ and
\begin{eqnarray*}
\left| \sum_{i=3}^{\infty} \frac {h_{\lambda}^{i-3} \left( b_{n}; x \right)}{i!} \right| <
\exp \left[ h_{\lambda} \left(b_{n}; x \right) \right] \to 1
\end{eqnarray*}
as $n\to \infty$.
By using Lemma \ref{lem2} once again, we have
\begin{eqnarray*}
&&
b_{n}^{2}\Big[ b_{n}^{2}\left( F_{\lambda}^{n}\left( a_{n}x+b_{n} \right) -\Lambda(x) \right) -
\kappa(x)\Lambda(x)\Big]
\\
&=&
b_{n}^{2}\Big[ b_{n}^{2}\left( \exp \left( h_{\lambda} \left(b_{n}; x\right) \right) - 1  \right) - \kappa(x)\Big] \Lambda(x)
\\
&=&
\left[ b_{n}^{2}\left( b_{n}^{2}h_{\lambda} \left(b_{n}; x\right) -\kappa(x) \right) +
b_{n}^{4} h_{\lambda}^{2} \left(b_{n}; x \right) \left( \frac {1}{2} + h_{\lambda} \left(b_{n}; x \right)
\sum_{i=3}^{\infty} \frac {h_{\lambda}^{i-3} \left(b_{n}; x\right)}{i!} \right)\right] \Lambda(x)
\\
&\to&
\left[ \omega(x) + \frac {\kappa^{2}(x)}{2} \right] \Lambda(x).
\end{eqnarray*}
The desired result follows.
\qed

Theorem \ref{thm2.2} establishes the asymptotic expansion of
$F^{n}_{\lambda}(a_{n}x+b_{n})$.
Meanwhile the
convergence rate of $F^{n}_{\lambda}(a_{n}x+b_{n})$ to its limit
distribution $\Lambda(x)$ is proportional to $1/\log n$ by Theorem
\ref{thm2.2} since one can check that $1/b_{n}^{2} = O\left(1/\log n\right)$ through \eqref{eq1.12}.

\vspace{1cm}

\noindent
{\bf Acknowledgements}~~This work was supported by the
National Natural Science Foundation of China no.11171275, the
Natural Science Foundation Project of CQ no. cstc2012jjA00029 and
the SWU grant for Statistics Ph.D.


\begin{thebibliography}{999}

\bibitem{Azzalini1}
Azzalini, A. (1985).
A class of distributions which includes the
normal ones.
{\it Scandinavian Journal of Statistics}, {\bf 12}, 171-178.



\bibitem{}
Bandyopadhyay, D., Lachos, V. H., Castro, L. M. and Dey, D. K. (2012).
Skew-normal/independent linear mixed models for censored responses with applications to HIV viral loads.
{\it Biometrical Journal}, {\bf 54}, 405-425.



\bibitem{Castro}
Castro, L. C. E. (1987).
Uniform rate of convergence in extreme-value theory: Normal and gamma models.
{\it Annales Scientifiques de l'Univesit\'{e} de Clerment-Ferrand 2, tome 90,
s\'{e}rie Probabilit\'{e}s et Applications}, {\bf 6}, 25-41.


\bibitem{Chang}
Chang, S. and Genton, M. G. (2007).
Extreme value distributions for the skew-symmetric family of distributions.
{\it Communications in Statistics---Theory and Methods}, {\bf 36}, 1705-1717.


\bibitem{}
Ghizzoni, T., Roth, G. and Rudari, R. (2012).
Multisite flooding hazard assessment in the Upper Mississippi River.
{\it Journal of Hydrology}, {\bf 412}, 101-113.



\bibitem{Goldie}
Goldie, C. M. and Resnick, S. I. (1988).
Distributions that are both
subexponential and in the domain of attraction of an extreme-value distribution.
{\it Journal of Applied Probability}, {\bf 20}, 706-718.

\bibitem{Hall}
Hall, P. (1979).
On the rate of convergence of normal extremes.
{\it Journal of Applied Probability}, {\bf 16}, 433-439.


\bibitem{}
Islam, A. E. and Alam, M. A. (2011).
Analyzing the distribution of threshold voltage degradation in nanoscale
transistors by using reaction-diffusion and percolation theory.
{\it Journal of Computational Electronics}, {\bf 10}, 341-351.


\bibitem{Leadbetter}
Leadbetter, M. R., Lindgren, G. and Rootz\'{e}n, H. (1983).
{\it Extremes and Related Properties of Random Sequences and Processes}.
Springer Verlag, New York.

\bibitem{Liao}
Liao, X. and Peng, Z. (2012).
Convergence rates of limit
distribution of maxima of lognormal samples.
{\it Journal of Mathematical Analysis and Applications}, {\bf 395}, 643-653.


\bibitem{}
Liu, X. -H., Li, Y. -G., Shen, G. -C., Weng, D. -M., Zhang, H. -W. and Zhang, F. -G. (2011).
Population structure of Schima superba in Qingliangfeng National Nature Reserve.
{\it Forest Research}, {\bf 24}, 28-32.

\bibitem{}
Mansourian, M., Kazemnejad, A., Kazemi, I., Zayeri, F. and Soheilian, M. (2012).
Bayesian analysis of longitudinal ordered data with flexible random effects using McMC:
application to diabetic macular Edema data.
{\it Journal of Applied Statistics}, {\bf 39}, 1087-1100.



\bibitem{Mills}
Mills, J. P. (1926).
Table of the ratio: area to bounding ordinate,
for any portion of the normal curve.
{\it Biometrika}, {\bf 18}, 395-400.

\bibitem{Mitrinovic}
Mitrinovi\'{c}, D. S. and Vasi\'{c}, P. M. (1970).
{\it Analytic Inequalities}.
Springer Verlag, New York.



\bibitem{Nair}
Nair, K. A. (1981).
Asymptotic distribution and moments of normal extremes.
{\it Annals of Probability}, {\bf 9}, 150-153.


\bibitem{}
Paulson, K. and Al-Mreri, A. (2011).
A rain height model to predict fading due to wet snow on terrestrial links.
{\it Radio Science},  {\bf 46}, Article Number: RS4010.



\bibitem{Peng}
Peng, Z., Nadarajah, S. and Lin, F. (2010).
Convergence rate of extremes from general error distribution.
{\it Journal of Applied Probability}, {\bf 47}, 668-679.


\bibitem{}
Pinheiro, M. and Esteves, P. S. (2012).
On the uncertainty and risks of macroeconomic forecasts: combining judgements with sample and model information.
{\it Empirical Economics}, {\bf 42}, 639-665.



\bibitem{Resnick}
Resnick, S. I. (1987).
{\it Extreme Values, Regular Variation and Point Processes}.
Springer Verlag, New York.


\bibitem{}
Saez, A. J., Prieto, F. and Sarabia, J. M. (2012).
A two-tail version of the PPS distribution with application to current account balance data.
{\it Physica A---Statistical Mechanics and Its Applications}, {\bf 391}, 5160-5171.




\bibitem{}
Siebert, A. B. and Ward, M. N. (2011).
Future occurrence of threshold-crossing seasonal rainfall totals: Methodology and application to sites in Africa.
{\it Journal of Applied Meteorology and Climatology},  {\bf 50}, 560-578.


\bibitem{}
Temko, A., Stevenson, N., Marnane, W., Boylan, G. and Lightbody, G. (2012).
Inclusion of temporal priors for automated neonatal EEG classification.
{\it Journal of Neural Engineering}, {\bf 9},  Article Number: 046002.



\end{thebibliography}
\end{document}